\documentclass[aps,prl,reprint,superscriptaddress]{revtex4-1}
\usepackage{graphicx}
\usepackage{bbm}

\newcommand{\bra}[1]{\langle #1|}
\newcommand{\ket}[1]{|#1\rangle}

\begin{document}

\title{Difficulty of distinguishing product states locally}
\author{Sarah Croke}
\email{sarah.croke@glasgow.ac.uk}
\author{Stephen M. Barnett}
\address{School of Physics and Astronomy, University of Glasgow, Glasgow G12 8QQ, UK}

\begin{abstract}
Non-locality without entanglement is a rather counter-intuitive phenomenon in which information may be encoded entirely in product (unentangled) states of composite quantum systems in such a way that local measurement of the subsystems is not enough for optimal decoding. For simple examples of pure product states, the gap in performance is known to be rather small when arbitrary local strategies are allowed. Here we restrict to local strategies readily achievable with current technology; those requiring neither a quantum memory nor joint operations. We show that, even for measurements on pure product states there can be a large gap between such strategies and theoretically optimal performance. Thus even in the absence of entanglement physically realizable local strategies can be far from optimal for extracting quantum information.
\end{abstract}

\maketitle

A composite quantum system is more than the sum of its parts; it can have properties that cannot be explained as a result of properties of its constituent particles. This may be thought of as a result of the superposition principle, and has far-reaching consequences, giving quantum theory a far richer structure than classical probability theory. Indeed entangled states, characterized by correlations between particles rather than local properties of constituent particles, famously exhibit correlations that cannot be reproduced by any local classical theory \cite{Bell}. In addition, there exist properties of composite systems that are not accessible through only local measurement of the subsystems: for example, a measurement which distinguishes between the $j=0$ and $j=1$ subspaces of a two spin-half system cannot be performed through only local measurements on the spins.

In contrast to classical probability theory therefore, local measurements together with post-processing of measurement results is not enough to perform all measurements allowed by quantum theory. In fact, due to the existence of incompatible observables, even classical communication \emph{between} subsystems (allowing the choice of measurement on one system to depend on the outcome of measurement on the other) gives the measuring parties new capabilities. In general, each of one-way classical communication, two-way classical communication, and quantum communication (the ability to send quantum states) is more powerful than the last \cite{Peres91,Bennett99,Chitambar13,Massar95,Brody96,Walgate00,Ghosh01,Groisman01,Walgate02,DiVincenzo02,Chen03,
Chen04,Rinaldis04,Owari08,Feng09,Calsamiglia10,Cohen11,Higgins11,Childs13,Nathanson13,Mahler13,Zhang14,Chitambar14,Chitambar14a,
Zhang15,Wang16}. Remarkably, this is true even for measurements on unentangled states, a phenomenon known as ``non-locality without entanglement'' \cite{Peres91,Bennett99,Childs13,Chitambar13}. In practical terms however, for measurements on product states, the known bounds on the gaps in performance are really rather small \cite{Childs13,Chitambar13}.

In this paper we introduce a new piece in the puzzle of local detection of quantum information, and show that two-way classical communication can significantly improve the distinguishability of pure, orthonormal product states. Such sets of states have two important features: they can be prepared in separated laboratories without either classical or quantum communication between labs, and they are perfectly distinguishable through a joint measurement. We derive optimal one-way strategies for two examples of such sets, one of which may be perfectly discriminated with only two rounds of classical communication, while the other, the so-called domino states, requires quantum communication for perfect discrimination. We show that any one-way scheme succeeds with probability at most $\simeq 85\%$ and $\simeq 84 \%$ respectively, a significant deficit compared to the globally optimal schemes.

Physically, current experimental capabilities are such that sequential measurement (i.e. one-way) strategies may be readily implemented in a variety of physical systems. Additional rounds of measurement and classical communication are more technologically challenging; they not only introduce additional errors and inefficiencies, but also require some sort of quantum memory in which to store the local systems in a many round protocol. Our examples demonstrate that even for the simplest cases of discriminating pure, orthonormal, unentangled states, quantum memories or joint control can be necessary for optimal, or even close-to-optimal performance.

We begin with a simple example, first considered by Groisman and Vaidman \cite{Groisman01}, which demonstrates the asymmetry of classical communication as a resource for performing joint measurements, and serves to illustrate some features of optimal measurement strategies we will need later. Consider the product basis of two-qubit states:
\begin{equation}
\begin{array}{rclrcl}
\ket{\psi_{00}} &=& \ket{0}_A \otimes \ket{0}_B, & \ket{\psi_{10}} &=& \ket{1}_A \otimes \ket{0+1}_B,\\
\ket{\psi_{01}} &=& \ket{0}_A \otimes \ket{1}_B, & \ket{\psi_{11}} &=& \ket{1}_A \otimes \ket{0-1}_B.
\end{array}
\end{equation}
where $\{ \ket{0}, \ket{1}\}$ form an orthonormal basis, by convention the eigenbasis of the Pauli operator $\sigma_z$, and $\ket{0 \pm 1} = \frac{1}{\sqrt{2}} \left( \ket{0} \pm \ket{1} \right)$ are the eigenstates of $\sigma_x$. Clearly, the states are perfectly distinguishable given one-way communication from Alice to Bob, but not the other way around: if Alice can send a message to Bob, she simply needs to measure in the $\{ \ket{0}, \ket{1} \}$ basis, and send the result of her measurement to Bob. Given result ``0'', Bob measures in the $\{ \ket{0}, \ket{1} \}$ basis, while given result ``1'' he measures in the $\{ \ket{0 \pm 1} \}$ basis. On the other hand, if Bob can send messages to Alice but not vice-versa, there is clearly no procedure that allows Alice and Bob to perfectly distinguish the states: Bob's local states are eigenstates of incompatible observables $\sigma_z$ and $\sigma_x$, and any measurement giving information about one must necessarily disturb the other, thus destroying the orthogonality of at least one pair of states.

It is instructive to consider what is the best Alice and Bob can do if they are limited to communication only from Bob to Alice. Note that Alice's system is essentially classical in that regardless of the information obtained from Bob, the only sensible measurement she can make is in the $z$-basis, which allows her to determine perfectly the index $i$ in the labelling $\{ \ket{\psi_{ij}} \}$, but gives no information about the index $j$. The role of Bob's measurement therefore is to provide the information which allows Alice to distinguish between the states within the pair $\{ \ket{\psi_{i0}}, \ket{\psi_{i1}} \}$, for each possible outcome $i$ of Alice's measurement. It follows that Bob's measurement must rule out as well as possible one state from each pair, leaving Alice with two remaining allowed states, which are perfectly distinguishable through Alice's measurement. There are four subsets of two states with the property that the states are perfectly distinguishable on Alice's side:
\begin{eqnarray*}
\mathcal{S}_{00} = \{ \ket{\psi_{00}}, \ket{\psi_{10}} \}, &\quad& \mathcal{S}_{01} = \{ \ket{\psi_{00}}, \ket{\psi_{11}} \}, \\
\mathcal{S}_{10} = \{ \ket{\psi_{01}}, \ket{\psi_{10}} \}, && \mathcal{S}_{11} = \{ \ket{\psi_{01}}, \ket{\psi_{11}} \}.
\end{eqnarray*}
Thus Alice and Bob's strategy, if classical communication is allowed from Bob to Alice only, is for Bob's measurement to optimally assign the state to one of these subsets, while Alice's measurement discriminates between the remaining two states within a subset.

If the states are equiprobable the measurement on Bob's system maximising the probability of correctly identifying the state is in fact quite well-known, as these states arise in the BB84 protocol for quantum key distribution \cite{BB84}. The index $i$ corresponds to the sender's choice of basis, while $j$ denotes the bit value. Naturally enough due to the symmetry of the set, the optimal measurement is in a basis intermediate to the $x$-basis and the $z$-basis: the so-called Breidbart basis \cite{Breidbart}:
\begin{eqnarray}
\ket{\phi_0} &=& \cos \frac{\pi}{8} \ket{0} + \sin \frac{\pi}{8} \ket {1}, \nonumber \\
\ket{\phi_1} &=& \sin \frac{\pi}{8} \ket{0} - \cos \frac{\pi}{8} \ket{1}. \label{breidbart}
\end{eqnarray}
Outcome 0 leads Bob to guess that the state belongs to subset $\mathcal{S}_{00}$ while outcome 1 leads to a guess of $\mathcal{S}_{11}$. In fact the optimal measurement is degenerate: measurement of either of the spin observables $\frac{1}{\sqrt2} (\sigma_z \pm \sigma_x)$ results in an optimal strategy. This succeeds with probability $\cos^2 \frac{\pi}{8} =\frac{1}{2} (1+\frac{1}{\sqrt2}) \simeq 0.85$, a significant deficit compared to the unit probability of success achievable when one-way communication is allowed from Alice to Bob \cite{Groisman01}.

Using this simple set as a building block we can construct our first example for which two-way classical communication provides a significant advantage over one-way classical communication for discriminating unentangled states. Consider the following orthonormal product basis of a $2 \otimes 4$ level system (introduced also in \cite{Feng09}):
\begin{equation}
\begin{array}{rclrcl}
\ket{\psi_{00}} &=& \ket{0}_A \ket{0}_B, & \ket{\psi_{02}} &=& \ket{0+1}_A \ket{2}_B,\\
\ket{\psi_{01}} &=& \ket{0}_A \ket{1}_B, & \ket{\psi_{03}} &=& \ket{0+1}_A \ket{3}_B,\\
\ket{\psi_{10}} &=& \ket{1}_A \ket{0+1}_B, & \ket{\psi_{12}} &=& \ket{0-1}_A \ket{2+3}_B, \\
\ket{\psi_{11}} &=& \ket{1}_A \ket{0-1}_B, & \ket{\psi_{13}} &=& \ket{0-1}_A \ket{2-3}_B.
\end{array}
\end{equation}
These states are perfectly distinguishable given just two rounds of classical communication, while any one-way scheme succeeds with probability at most $\cos^2 \frac{\pi}{8}$, as we now show.

We first note that Bob can learn in which subspace $\{ \ket{0}, \ket{1} \}$ or $\{ \ket{2}, \ket{3} \}$ his state lies without disturbing any of the states, via a von Neumann measurement with projectors $\{ \ket{0} \bra{0} + \ket{1} \bra{1}, \ket{2} \bra{2} + \ket{3} \bra{3} \}$, thus learning to which of the two subsets $\mathcal{S}_0 = \{ \ket{\psi_{00}}, \ket{\psi_{01}}, \ket{\psi_{10}}, \ket{\psi_{11}} \}$ or $\mathcal{S}_1 = \{ \ket{\psi_{02}}, \ket{\psi_{03}}, \ket{\psi_{12}}, \ket{\psi_{13}} \}$ the shared state belongs. Each subset is equivalent (up to local unitaries) to the simpler set discussed above. This observation simplifies the analysis of optimal schemes.

There are three cases of interest: i) Two-way classical communication: Bob measures first, and tells Alice to which of the subsets $\mathcal{S}_0$ or $\mathcal{S}_1$ the state belongs. Each subset is perfectly distinguishable with one-way communication from Alice to Bob, so just one more round of communication is needed for perfect discrimination. ii) One-way communication from Alice to Bob: within each subset, the shared states are perfectly distinguishable with one-way measurement from Alice to Bob. However, Alice doesn't know in which subset the state lies. Clearly Alice's measurement must simultaneously distinguish, as well as possible, between the states $\{ \ket{0}, \ket{1} \}$ and $\{ \ket{0+1}, \ket{0-1} \}$. But this is simply the same problem as we have seen previously, and an optimal measurement for Alice is again in the Breidbart basis eqn. (\ref{breidbart}). Alice communicates the result of measurement to Bob; given outcome 0, Bob's measurement discriminates perfectly between the states $\{ \ket{0}, \ket{1} \}$ and $\{\ket{2}, \ket{3} \}$, given outcome 1 he instead measures in the basis $\{ \ket{0 \pm 1}, \ket{2 \pm 3} \}$. The combination of Alice's and Bob's measurement results leads to a unique guess as to the state, which is correct with probability $\cos^2 \frac{\pi}{8}$. iii) One-way communication from Bob to Alice: Bob knows in which subset $\mathcal{S}_0$, $\mathcal{S}_1$ the state lies. In each case as we have noted, the resulting set is equivalent to our simple example discussed above, for which the optimal one-way strategy succeeds with probability $\cos^2 \frac{\pi}{8}$. Thus any one-way scheme, regardless of the direction of communication, succeeds with probability at most $\cos^2 \frac{\pi}{8} \simeq 85 \%$, while two rounds of classical communication are sufficient to discriminate the states perfectly.

\begin{figure}[h!]
\begin{center}
\includegraphics[width=60mm]{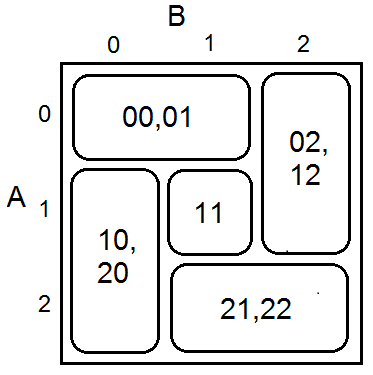}
\end{center}
\caption{Graphical representation of the domino states.}
\label{DominoFig}
\end{figure}

For our second example, we turn to the domino states \cite{Bennett99}, an orthonormal basis of a $3 \otimes 3$ level system given by:
\begin{equation}
\begin{array}{c}
\begin{array}{rclrcl}
\ket{\psi_{00}} &=& \ket{0} \ket{0-1}, & \ket{\psi_{10}} &=& \ket{1+2} \ket{0}, \\
\ket{\psi_{01}} &=& \ket{0} \ket{0+1}, & \ket{\psi_{11}} &=& \ket{1} \ket{1},  \\
\ket{\psi_{02}} &=& \ket{0-1} \ket{2},  & \ket{\psi_{12}} &=& \ket{0+1} \ket{2},
\end{array} \\
\begin{array}{rcl}
\ket{\psi_{20}} &=& \ket{1-2} \ket{0}, \\
\ket{\psi_{21}} &=& \ket{2} \ket{1-2}, \\
\ket{\psi_{22}} &=& \ket{2} \ket{1+2},
\end{array}
\end{array}
\end{equation}
A useful graphical representation of these states is given in Fig. \ref{DominoFig}.
This example is of a different flavour to the previous one, as the states are globally perfectly distinguishable, but two-way classical communication, even in the limit of infinite rounds of measurement and communication, is not enough to perfectly distinguish the states \cite{Bennett99}. The known upper bound on the probability of error of any scheme with two-way communication however is so small as to be negligible for all practical purposes: $1.9 \times 10^{-8}$ \cite{Childs13}. We find, by contrast, that the best one-way strategy has an error of more than $16 \%$.

We begin by simply stating the optimal sequential measurement, which is somewhat intuitive, and give some of the technical details later. We assume the states are equiprobable, and we suppose that $A$ is measured first, which due to the symmetry of the states we can do without any loss of generality. There are 8 subsets of the set of domino states which are perfectly distinguishable on system $B$ alone, these are:
\begin{equation}
\begin{array}{rclrcl}
\mathcal{S}_0 &=& \{ \ket{\psi_{00}}, \ket{\psi_{01}}, \ket{\psi_{02}} \}, & \mathcal{S}_4 &=& \{ \ket{\psi_{10}}, \ket{\psi_{11}}, \ket{\psi_{02}} \}, \\
\mathcal{S}_1 &=& \{ \ket{\psi_{00}}, \ket{\psi_{01}}, \ket{\psi_{12}} \}, & \mathcal{S}_5 &=& \{ \ket{\psi_{10}}, \ket{\psi_{11}}, \ket{\psi_{12}} \}, \\
\mathcal{S}_2 &=& \{ \ket{\psi_{10}}, \ket{\psi_{21}}, \ket{\psi_{22}} \}, & \mathcal{S}_6 &=& \{ \ket{\psi_{20}}, \ket{\psi_{11}}, \ket{\psi_{02}} \}, \\
\mathcal{S}_3 &=& \{ \ket{\psi_{20}}, \ket{\psi_{21}}, \ket{\psi_{22}} \}, & \mathcal{S}_7 &=& \{ \ket{\psi_{20}}, \ket{\psi_{11}}, \ket{\psi_{12}} \}.
\end{array} \label{dominosubsets}
\end{equation}
The optimal sequential strategy assigns the state to one of these subsets in the first step, and discriminates perfectly the states within the subset in the second step. This succeeds with probability $\simeq 83.6 \%$, as we show later.

Proving that this is indeed optimal is less straight-forward than the previous cases. Our strategy is to place an upper bound on the probability of success of any sequential strategy, by considering a simpler, related discrimination problem, and then show that this bound is achievable. Thus, to bound the probability of success in identifying the state, rather than trying to discriminate between all nine states, let us suppose instead that we simply try to determine to which of the three subsets $\{ \ket{\psi_{00}}, \ket{\psi_{01}}, \ket{\psi_{02}}\}$, $\{ \ket{\psi_{10}}, \ket{\psi_{11}}, \ket{\psi_{12}}\}$, or $\{ \ket{\psi_{20}}, \ket{\psi_{21}}, \ket{\psi_{22}}\}$ the state belongs; that is, with our choice of labeling $\ket{\psi_{jk}}$, we try to determine the index $j$, without worrying about $k$. This is equivalent to discriminating between the equiprobable mixed states
\begin{equation}
\rho_j = \frac{1}{3} \sum_k \ket{\psi_{jk}} \bra{\psi_{jk}}.
\label{rho}
\end{equation}
This problem of subset discrimination is strictly easier than our original problem: any measurement to discriminate between all nine domino states may also be used for subset discrimination, and performs at least as well for this task. Thus the success probability of the optimal sequential measurement discriminating the states $\{\rho_i \}$ is at least as high as the probability of success of any sequential measurement discriminating all nine domino states:
$$
P_{\rm corr}^{\rm seq} (\{ \psi_i \}) \leq P_{\rm corr}^{\rm seq} (\{ \rho_i \}).
$$
This set further has the rather nice property that Bob's system is essentially classical; his state is always diagonal in the $\{ \ket{0}, \ket{1}, \ket{2} \}$ basis, thus regardless of the measurement on $A$, the only sensible measurement on $B$ is in this basis. The role of Alice's measurement then, as in our first example, is simply to inform Bob how to interpret his measurement result. By inspection of the states $\rho_j$ (eq. \ref{rho}) we see that if Bob obtains outcome 0, Alice's measurement must discriminate between the orthonormal states $\{ \ket{0}, \ket{1+2}, \ket{1-2} \}$, to provide Bob with the information required to identify one of the states $\{ \rho_0, \rho_1, \rho_2 \}$ respectively. Similar observations for Bob's outcomes 1 and 2 reveal that the role of Alice's measurement is to distinguish simultaneously as well as possible the states within the three bases $\{ \ket{0}, \ket{1 \pm 2} \}$, $\{ \ket{0}, \ket{1}, \ket{2} \}$, $\{ \ket{0 \pm 1}, \ket{2} \}$. There are 27 subsets containing exactly one state from each basis: Alice's job is to optimally discriminate between these subsets. This sounds like rather a daunting task, however it turns out, as we discuss below, that the optimal strategy is one that only ever identifies those subsets $\{ \mathcal{S}_0, \cdots, \mathcal{S}_7 \}$ given in eqn. (\ref{dominosubsets}).

Thus Alice's measurement giving the optimal one-way strategy for this simpler problem is precisely that conjectured to be optimal for discriminating all nine domino states. This succeeds with probability $83.6\%$, which as argued above is therefore an upper bound on the success probability for discriminating the domino states with only one-way classical communication. Further, this bound is achievable via the conjectured sequential scheme given above, which we therefore conclude is an optimal one-way strategy. In the remainder of the paper we discuss some of the technical details of the derivation of Alice's optimal measurement. We give more details and an alternative proof elsewhere \cite{Croke16}.

Alice wishes to optimally assign the state to one of the subsets $\{ \mathcal{S}_i \}$. She performs a measurement on her system, and upon obtaining outcome $j$ takes this to indicate that the state belonged to the subset $\mathcal{S}_j$. The most general measurement she can perform on her system is described by a probability operator measure (POM) \cite{SteveBook}, also known as a positive operator-valued measure (POVM) \cite{Peres95}, that is a set of positive operators $\{ \pi_j \}$ that form a resolution of the identity
\begin{equation}
\pi_j  \geq  0, \qquad \sum_j \pi_j = I. \label{POVM}
\end{equation}
The probability of obtaining outcome $j$ in a measurement on a system prepared in state $\rho$ is given by the Born rule: $P(j|\rho) = {\rm Tr}(\rho \pi_j)$. Thus, the probability that Alice chooses a subset containing the state prepared is given by:
\begin{eqnarray*}
{\rm P}_{corr} &=& \sum_{ij} \frac{1}{9} \sum_{k|\ket{\psi_{ij}} \in \mathcal{S}_k} {\rm Tr}_{AB}( \ket{\psi_{ij}} \bra{\psi_{ij}} \pi_k ) \\
&=& \frac{8}{3} \sum_k \frac{1}{8} {\rm Tr}_A \left( \sigma_k \pi_k \right)
\end{eqnarray*}
where in the last line we have defined $\sigma_k =  {\rm Tr}_B \left(\frac{1}{3} \sum_{ij|\ket{\psi_{ij}} \in \mathcal{S}_k} \ket{\psi_{ij}} \bra{\psi_{ij}} \right)$. $\sigma_k$ is thus the density operator obtained by taking an equal mixture of the states in the subset $\mathcal{S}_k$, traced over Bob's system. It follows from the above that $\{ \pi_k \}$ is the optimal measurement to discriminate the equiprobable states $\{ \sigma_k \}$. Explicitly, the states $\sigma_k$ are given by:
\begin{eqnarray*}
\sigma_0 &=& \frac{2}{3} \ket{0} \bra{0} + \frac{1}{3} \ket{0-1} \bra{0-1}, \\
\sigma_1 &=& U \sigma_0 U^\dagger, \quad \sigma_2 = V U \sigma_0 U^\dagger V^\dagger, \quad \sigma_3 = V \sigma_0 V^\dagger, \\
\sigma_4 &=& \frac{1}{3} \ket{0-1} \bra{0-1} + \frac{1}{3} \ket{1} \bra{1} + \frac{1}{3} \ket{1+2} \bra{1+2}, \\
\sigma_5 &=& U \sigma_4 U^\dagger, \quad \sigma_6 = U V \sigma_4 V^\dagger U^\dagger, \quad \sigma_7 = V \sigma_4 V^\dagger,
\end{eqnarray*}
where we have introduced the unitary operators:
\begin{eqnarray*}
U &=& - \ket{0} \bra{0} + \ket{1} \bra{1} + \ket{2} \bra{2}, \\
V &=& \ket{0} \bra{2} + \ket{1} \bra{1} + \ket{2} \bra{0}.
\end{eqnarray*}
Any measurement $\{ \pi_j \}$ discriminating optimally between the states $\{\rho_j \}$ with priors $\{ p_j \}$ satisfies the so-called Helstrom conditions, which are known to be both necessary and sufficient \cite{Holevo73,Yuen75,Helstrom76,Barnett09}:
\begin{eqnarray}
\sum_j p_j \rho_j \pi_j - p_k \rho_k & \geq & 0, \label{hel1} \\
\pi_i (p_i \rho_i - p_j \rho_j ) \pi_j &=& 0. \label{hel2}
\end{eqnarray}
Denoting $\Gamma = \sum_i p_i \rho_i \pi_i$, an alternative, equivalent condition, which is sometimes easier to use in practice, is obtained by summing over $i$ in eqn. \ref{hel2}, giving
\begin{equation}
\left( \Gamma - p_j \rho_j \right) \pi_j = 0. \label{hel3}
\end{equation}

Our strategy to find the optimal probability of success for Alice's measurement is to use the Helstrom conditions constructively to find $\Gamma = \frac{1}{8} \sum_i \sigma_i \pi_i$ such that the operators $\Gamma - \frac{1}{8} \sigma_j$ are rank-two positive semi-definite for all $j$. Thus eqn \ref{hel1} is satisfied, and choosing $\pi_j$ to be a weighted projector onto the zero eigenvalue eigenstate of $\Gamma - \frac{1}{8} \sigma_j$ ensures that eqn \ref{hel3} is also satisfied. If we can choose the weights such that the resulting operators $\pi_j$ sum to the identity, we have succeeded in finding an optimal measurement. This task is facilitated by noting that the states have rather a lot of symmetry. It is clear that the set $\{ \sigma_k \}$ is invariant under the action of the two unitary operators $U$ and $V$. Hence we search for a measurement $\{ \pi_i \}$ with the same symmetry properties. It follows that we are searching for an operator $\Gamma = \frac{1}{8} \sum_i \sigma_i \pi_i$ which is invariant under $U$ and $V$: $\Gamma = U \Gamma U^ \dagger = V \Gamma V^\dagger$, which further implies that $\Gamma$ is of the form:
$$
\Gamma = p \left( \ket{0} \bra{0} + \ket{2} \bra{2} \right) + q \ket{1} \bra{1}
$$
Finally, using the symmetry again, we need only check the condition (\ref{hel1}) for $\sigma_0$ and $\sigma_4$, the rest follow by construction. Imposing that each of $\Gamma - \frac{1}{8} \sigma_0$ and $\Gamma - \frac{1}{8} \sigma_4$ have one zero eigenvalue allows us to solve for $p$ and $q$, giving $p \simeq 0.110$, $q \simeq 0.093$, and $P_{\rm corr} = \frac{8}{3} {\rm Tr}(\Gamma) = \frac{8}{3} (2 p + q) \simeq 0.836$. We note without proof that we can indeed form a resolution of the identity from the resulting operators $\{ \pi_i \}$, which therefore define an optimal measurement. Finally we recall that the simplified problem of discriminating the three mixed states $\{ \rho_0, \rho_1, \rho_2 \}$ given above leads to a problem of discriminating between 27 subsets on Alice's side. Performing the same analysis as above it is readily verified that the measurement just derived, that optimally discriminates the 8 subsets $\{ \mathcal{S}_0, \cdots, \mathcal{S}_7 \}$ remains optimal. It is tedious but straight-forward to check that the Helstrom condition \ref{hel1} is satisfied for the remaining states, which are therefore never identified.

We thus find that for these cases, it is two-way classical communication that significantly boosts the distinguishability of the states and that quantum communication provides only a small additional boost in one case. It is known of course that two-way communication is more powerful than one-way communication; however explicit, quantitative gaps for the problem of discriminating orthogonal states have been shown in the literature only for measurements on entangled states (see e.g. \cite{Owari08}, \cite{Nathanson13}). It is surprising that even a single round of two-way classical communication can provide such a significant improvement in measurements on pure product states.

The optimal sequential measurement is a well-motivated indicator of achievable experimental performance in local measurement schemes: many rounds of measurement and classical communication quickly become infeasible, and further require quantum memories to store the local quantum systems. Thus, even for rather simple cases, and when information is encoded in product states, an appreciable gap can exist between the performance of the best readily achievable local measurement and the theoretically allowed optimum measurement. In practical terms this is arguably a much stronger manifestation of non-locality without entanglement than the known theoretical cases -- albeit of a different flavour.

\begin{acknowledgements}
This work was supported by the University of Glasgow College of Science and Engineering through an Academic Returners grant (S.C.) and by the Royal Society Research Professorships (S.M.B.).
\end{acknowledgements}

\end{document}